\documentstyle[preprint,aps]{revtex}
\preprint{USM-TH-147}
\begin{document}
\title{Static potential in a topologically massive Born-Infeld theory}
\author{ Patricio Gaete \thanks{E-mail: patricio.gaete@fis.utfsm.cl}}
\address{Departamento de F\'{\i}sica, Universidad T\'ecnica F.
Santa Mar\'{\i}a, Valpara\'{\i}so, Chile} \maketitle

\begin{abstract}
For a (2+1)-dimensional topologically massive Born-Infeld theory,
we compute the interaction potential within the structure of the
gauge-invariant but path-dependent variables formalism. The result
is equivalent to that of $QED_3$ with a Thirring interaction term
among fermions, in the short distance regime.
\end{abstract}
\smallskip

PACS number(s): 11.10.Ef, 11.10.Kk

\section{INTRODUCTION}

Topologically massive gauge theories in three-dimensional
spacetime have been studied by various authors in the last few
years\cite{Deser,Dunne,Khare}. An example of such a class of
theories is the Maxwell-Chern-Simons theory, which is endowed with
interesting features such as a massive gauge field and physical
excitations with fractional statistics (anyons) which interpolate
between bosons and fermions. In addition, they are interesting
because of its connection to the high-temperature limit of
four-dimensional theories\cite{Appelquist,Templeton} as well as
for their applications to condensed matter physics \cite{Stone}.

We further observe that recently a great deal of attention has
been devoted to the study of non-linear electrodynamics
(Born-Infeld theory) due to its arising naturally in D-brane
physics\cite{Tseytlin,Gibbons}. Let us also recall here that Born
and Infeld\cite{Born} suggested to modify Maxwell's
electromagnetism with the goal of avoiding divergences such as the
infinite self-energy of a point charge.

On the other hand, the study of duality symmetry in gauge theories
has also attracted considerable attention in order to provide an
equivalent description of physical phenomena by distinct theories.
As is well known, duality refers to an equivalence relation
between two or more quantum field theories whose corresponding
classical theories are different. An interesting and illustrative
example on this subject arises when one considers three dimensions
of spacetime, where it was shown in
Refs.\cite{Deser,Townsend,Jackiw} that the self-dual and
Maxwell-Chern-Simons theories are identical as quantum theories.
Within this context a $(2+1)$-dimensional topologically massive
Born-Infeld theory has been studied in Ref.\cite{Tripathy}.
However, in this Letter we wish to further elaborate on the
physical content of this theory. To this end we will compute the
lowest order modification of the static potential due to the
presence of the Born-Infeld term. In fact, we will show that the
static potential for the topologically massive Born-Infeld theory
agrees with that of the Maxwell-Chern-Simons theory with a
Thirring interaction term among fermions, in the short distance
regime. In this way we establish a new connection between both
theories, in the hope that this will be helpful to understand
better effective gauge theories in $(2+1)$ dimensions. Our
calculation is based on the gauge-invariant but path-dependent
variables formalism. According to this formalism, the interaction
potential between two static charges is obtained once a judicious
identification of the physical degrees of freedom is
made\cite{Gaete}. This methodology, in our view, is of interest
both for its simplicity and physical content.

\section{INTERACTION ENERGY}

As already stated, our objective is to compute explicitly the
interaction energy between static pointlike sources for
topologically massive Born-Infeld theory. To this end we shall
first carry out its Hamiltonian analysis. The starting point is
the Lagrangian \cite{Tripathy}:
\begin{equation}
{\cal L} = \beta ^2 \left\{ {1 - \sqrt {1 + \frac{1}{{2\beta ^2
}}F_{\mu \nu } F^{\mu \nu } } } \right\} + \frac{\theta
}{4}\varepsilon ^{\mu \nu \lambda } A_\mu  F_{\nu \lambda }  - A_0
J^0, \label{bics1}
\end{equation}
where $J^0$ is the external current. The parameter $ \beta $
measures the nonlinearity of the theory and in the limit $ \beta
\to \infty $ the Lagrangian (\ref{bics1}) reduces to the
Maxwell-Chern-Simons theory. In order to handle the square root in
the Lagrangian (\ref{bics1}), we introduce an auxiliary field $v$,
such that its equation of motion gives back the original
theory\cite{Tseytlin}. This allows us to rewrite the Lagrangian
 (\ref{bics1}) in the form
\begin{equation}
{\cal L} = \beta ^2 \left\{ {1 - \frac{v}{2}\left( {1 +
\frac{1}{{2\beta ^2 }}F_{\mu \nu } F^{\mu \nu } } \right) -
\frac{1}{{2v}}} \right\} + \frac{\theta }{4}\varepsilon ^{\mu \nu
\lambda } A_\mu  F_{\nu \lambda }  - A_0 J^0. \label{bics2}
\end{equation}
In order to derive the canonical Hamiltonian, we note that the
canonical momenta are $ \Pi ^\mu = vF^{\mu 0} + \frac{\theta}{2}
\varepsilon ^{0\mu \nu } A_\nu$.  This yields the usual primary
constraint $\Pi ^0=0$ and $ p \equiv \frac{{\partial {\cal
L}}}{{\partial {\dot v}}} = 0$. Standard techniques for
constrained systems then lead to the following canonical
Hamiltonian
\begin{equation}
\begin{array}{r}
H_C  = \int {d^2 } x\left\{ { - \frac{1}{{2v}}\left( {\Pi _i \Pi^i
+ \theta \varepsilon ^{ij} A_i \Pi _j  + \frac{{\theta ^2 }}{2}A_i
A^i } \right) + \frac{v}{2}\left( {\frac{1}{2}F_{ij} F^{ij}  +
\beta ^2 } \right) - \beta ^2 } \right\} \nonumber + \\ + \int
{d^2 } x\left\{ { - A_0 \left( {\partial _i \Pi ^i  + \frac{\theta
}{2}\varepsilon ^{ij} \partial _i A_j - J^0 } \right)} \right\}.
\\ \label{bics3}
\end{array}
\end{equation}
Temporal conservation of the primary constraint $\Pi ^0$ leads to
the secondary constraint $\Gamma_1(x)\equiv\partial _i \Pi ^i +
\frac{\theta }{2}\varepsilon ^{ij} \partial _i A_j  - J^0=0$. The
consistency condition for the $p$ constraint yields no further
constraints and just determines the field $v$,
\begin{equation}
v = \sqrt {\frac{{1 - \frac{1}{{\beta ^2 }}\left( {\Pi ^i \Pi _i +
\theta \varepsilon ^{ij} A_i \Pi _j  + \frac{{\theta ^2 }}{2}A^i
A_i } \right)}}{{1 + \frac{1}{{2\beta ^2 }}F^{ij} F_{ij} }}},
\label{bics4}
\end{equation}
which will be used to eliminate $v$. The extended Hamiltonian that
generates translations in time then reads $H = H_C  + \int d x
\left( {c_0 (x)\Pi_0 (x) + c_1 (x)\Gamma _1 (x)} \right)$, where
$c_0(x)$ and $c_1(x)$ are the Lagrange multipliers. Since $ \Pi_0
= 0$ for all time and $ \dot{A}_0 \left( x \right) = \left[ {A_0
\left( x \right),H} \right] = c_0 \left( x \right)$, which is
completely arbitrary, we discard $ A_0 \left( x \right)$ and $ \Pi
_0 \left( x \right)$ because they add nothing to the description
of the system. As a result, the Hamiltonian becomes
\begin{equation}
H  = \int {d^2 } x\left\{ {\beta ^2 \sqrt {\left( {1 + \frac{{B^2
}}{{\beta ^2 }}} \right)\left( {1 + \frac{{{\bf D}^2 }}{{\beta ^2
}}} \right)}  - \beta ^2  - c^ {\prime }  \left( x \right)\left(
{\partial _i \Pi ^i  + \theta \varepsilon ^{ij}
\partial _i A_j  - J^0 } \right)} \right\}
, \label{bics5}
\end{equation}
where $c^ \prime  \left( x \right) = c_1 \left( x \right) - A_0
\left( x \right)$, $B^2  = \frac{1}{2}F_{ij} F^{ij}$ and ${\bf
D}^2 = - \left( {\Pi _i \Pi ^i  + \theta \varepsilon ^{ij} A_i \Pi
_j + \frac{{\theta ^2 }}{2}A_i A^i } \right)$.

Along with the first class constraint $\Gamma_1(x)$ (Gauss' law)
we impose one gauge constraint such that the full set of
constraints becomes second class. A convenient choice is found to
be
\begin{equation}
\Omega _2 \left( x \right) \equiv \int\limits_{C_{\xi x} } {dz^\nu
} A_\nu  \left( z \right) = \int_0^1 {d\lambda } x^i A_i \left(
{\lambda x} \right) = 0, \label {bics6}
\end{equation}
where  $\lambda$  $\left( {0 \le \lambda  \le 1} \right)$ is the
parameter describing the spacelike straight path between the
reference points $ \xi ^k $ and $ x^k $ , on a fixed time slice.
For simplicity we have assumed the reference point $\xi^k=0$. The
choice (\ref{bics6}) leads to the Poincar\'{e} gauge
\cite{Gaete2}. According to the Dirac method, we arrive at the
following nonvanishing Dirac bracket:
\begin{equation}
\left\{ A_{i}(x),\pi ^{j}(y)\right\} ^{*}=g _{i}^{j}\delta
^{(2)}\left(
x-y\right) -\partial _{i}^{x}\int_{0}^{1}d\lambda \text{ }%
x^{j}\delta ^{(2)}\left( \lambda x-y\right) .  \label{bics7}
\end{equation}

Since we are interested in estimating the lowest-order correction
to the interaction energy, we will retain only the leading
quadratic term in the expression (\ref{bics5}). Thus the
Hamiltonian may be written as
\begin{equation}
H = \int {d^2 } x\left( {\frac{1}{2}{\bf D}^2  + \frac{1}{2}B^2 }
\right). \label{bics8}
\end{equation}
In order to make clear our subsequent work, we now write the Dirac
brackets in terms of $ B=\varepsilon _{ij}\partial ^{i}A^{j}$ and
$ D^i  = \Pi ^i  - \frac {\theta}{2} \varepsilon ^{ij} A_j$
fields, that is,
\begin{equation}
\left\{ {D_i \left(  x \right),D_j \left(  y \right)} \right\}^
* = -\theta \varepsilon _{ij}\delta ^{(2)} \left( x - y \right) ,
\label{bics9a}
\end{equation}
\begin{equation}
\left\{ {B \left(  x \right),B \left(  y \right)} \right\}^
* = 0 , \label{bics9b}
\end{equation}
\begin{equation}
\left\{ {D_i \left( x \right),B\left(  y \right)} \right\}^
* = - \varepsilon _{ij} \partial _x^j \delta ^{(2)} \left( x - y
\right). \label{bics9c}
\end{equation}
It gives rise to the following equations of motion for $D_i$ and
$B$ fields:
\begin{equation}
{\dot D}_i \left( x \right) =  - \theta \varepsilon _{ij} D_j
\left( x \right) + \varepsilon _{ij} \partial _j B\left( x \right)
,\label{bics10}
\end{equation}
\begin{equation}
{\dot B}\left(  x \right) =  - \varepsilon _{ij} \partial _i D_j
\left(  x \right). \label{bics11}
\end{equation}
In the same way, we write the Gauss law as:
\begin{equation}
\partial _i D_L^i  + \theta B  - J^0  =
0, \label{bics12}
\end{equation}
where $D_L^i$ refers to the longitudinal part of $D^i$. As a
consequence of Eqs. (\ref{bics10}) and  (\ref{bics11}), the static
fields are given by
\begin{equation}
B =  - \frac{{\theta J^0 }}{{\nabla ^2  - \theta ^2 }},
\label{bics13}
\end{equation}
\begin{equation}
D_i = \frac{1}{\theta }\partial _i B , \label{bics14}
\end{equation}
where $\nabla ^2$ is the two-dimensional Laplacian. For $ J^0
\left( {t,{\bf x}} \right) = e\delta ^ {(2)} \left( {\bf x}- {\bf
a} \right)$, expressions (\ref{bics13}) and (\ref{bics14})
immediately show that
\begin{equation}
B\left( x \right) = \frac{{e\theta }}{{2\pi }}K_0 \left( {\theta
|\bf x - \bf a|} \right), \label{bics15}
\end{equation}
\begin{equation}
D^i (x) =  - \frac{{e\theta }}{{2\pi }}\frac{{|{\bf x - \bf a}|^i
}}{{|{\bf x - \bf a}|}}K_1 \left( {\theta |{\bf x - \bf a}|}
\right), \label{bics16}
\end{equation}
where $K_0$ and $K_1$ are modified Bessel functions. However, it
follows from the above discussion that the electric field takes
the form
\begin{equation}
E^i \left( x \right) = \left\{ {1 - \frac{1}{2}\left(
{\frac{{e\theta }}{{2\pi \beta }}} \right)^2 \left[ {K_1^2 \left(
{\theta |{\bf x - \bf a}|} \right) - K_0^2 \left( {\theta |{\bf x
- \bf a}|} \right)} \right]} \right\}\partial ^i \left( { -
\frac{{J^0 }}{{\nabla ^2 - \theta ^2 }}} \right). \label{bics17}
\end{equation}

We pass now to the calculation of the potential energy for a pair
of static pointlike opposite charges at ${\bf y}$ and ${\bf
y}^\prime$. The procedure we shall follow is based on the
expression
\begin{equation}
V \equiv e\left( {{\cal A}_0 \left( \bf y \right) - {\cal A}_0
\left( {\bf y^ \prime} \right)} \right), \label{pot4}
\end{equation}
where the physical scalar potential ${\cal A}_0$ is expressed in
terms of the electric field
\begin{equation}
{\cal A}_0 \left( {t,\bf x} \right) = \int_0^1 {d\lambda } x^i E_i
\left( {t,\lambda \bf x} \right). \label{pot5}
\end{equation}
At this point it is worth to emphasize that Eq.(\ref{pot5})
follows from the vector gauge-invariant field \cite{Gaete}:
\begin{equation}
{\cal A}_\mu  \left( x \right) \equiv A_\mu  \left( x \right) +
\partial _\mu \left( { - \int_\xi ^x {dz^\mu  } A_\mu  \left( z
\right)} \right), \label{pot6}
\end{equation}
where, as in Eq.(\ref{bics6}), the line integral appearing in the
above expression is along a spacelike path from the point $\xi$ to
$x$, on a fixed time slice. At the same time, it should be noted
that the gauge-invariant variables (\ref{pot6}) commute with the
sole first class constraint (Gauss' law), supporting the fact that
these fields are physical variables \cite{Dirac}.

Having made these observations and using Eq.(\ref{pot5}), we can
write immediately the following expression for the physical scalar
potential:
\begin{equation}
{\cal A}_0 \left( {t,\bf x} \right) = \int_0^1 {d\lambda } \left\{
{1 - \frac{1}{2}\left( {\frac{{e\theta }}{{2\pi \beta }}}
\right)^2 \left[ {K_1^2 \left( {\theta |\lambda {\bf x} - {\bf
a}|} \right) - K_0^2 \left( {\theta |\lambda {\bf x} - {\bf a}|}
\right)} \right]} \right\}x^i \partial _i^{\lambda {\bf x}} \left(
{ - \frac{{J^0 \left( {\lambda {\bf x}} \right)}}{{\nabla
_{\lambda {\bf x}}^2 - \theta ^2 }}} \right), \label{pot7}
\end{equation}
where $J^0$ is the external current. Again, for $ J^0 \left(
{t,{\bf x}} \right) = e\delta ^ {(2)} \left( {\bf x}- {\bf a}
\right)$, Eq.(\ref{pot7}) becomes
\begin{equation}
{\cal A}_0 \left( {t,{\bf x}} \right) = \frac{e}{{2\pi }}\left(
{K_0 \left( {\theta |{\bf x} - {\bf a}|} \right) - K_0 \left(
{\theta |{\bf a}|} \right)} \right) - \frac{{e^3 }}{{32\pi ^3
\beta ^2 }}\left( {\frac{1}{{|{\bf x} - {\bf a}|^2 }} -
\frac{1}{{|{\bf a}|^2 }}} \right).   \label{pot8}
\end{equation}
Since we are only interested in estimating the lowest-order
modification of the static potential due to the presence of the
Born-Infeld term, we have dropped a logarithmic factor in
expression (\ref{pot8}) which does not contribute in the limit
$\beta\gg\theta$.

Accordingly, the potential for two opposite charges $e$ and $-e$
located at $\bf y$ and $\bf y \prime$ is then given by
\begin{equation}
V = \frac{{e^2 }}{\pi }\ln \left( {\theta |{\bf y} - {\bf y}^
{\prime} |} \right) - \frac{{e^4 }}{{16\pi ^3 \beta ^2
}}\frac{1}{{|{\bf y} - {\bf y}^ {\prime} |^2 }}, \label{pot9}
\end{equation}
in the short distance regime. It is interesting to note that this
is exactly the result obtained for $QED_3$ with a Thirring
interaction term among fermions, the so-called generalized
Maxwell-Chern-Simons gauge theory \cite{Gaete,Ghosh}. Thus, from a
physical point of view, one is led to the conclusion that the
effect of including the Born-Infeld term is to generate stable
bound states of quark-antiquark pairs at short distances, exactly
as it happens with the  Thirring term in the generalized
Maxwell-Chern-Simons theory. In summary, the above analysis
reveals that, although both theories are different, the physical
content is identical in the short distance regime.

\section{ACKNOWLEDGMENTS}

I would like to thank G. Cvetic for helpful comments on the
manuscript. I would also like to thank I. Schmidt for his support.

\end{document}